\documentstyle[preprint,aps]{revtex}
\begin{document}
\draft

\title{Neutron spin-dependent structure function, Bjorken sum rule,
and first evidence for singlet contribution at low $x$}
\author{J. Soffer \cite{1} and  O. V. Teryaev\cite{2}}
\address{Centre de Physique Th\'eorique - CNRS - Luminy,\\
Case 907 F-13288 Marseille Cedex 9 - France\\
and\\
Bogoliubov Laboratory of Theoretical Physics, \\
Joint Institute for Nuclear Research, Dubna, 141980, Russia}
\maketitle

\begin{abstract}
We perform the isospin decomposition of proton and neutron SLAC data
for moderately low $x$ in the region $0.01 \leq x \leq 0.1$.
The isovector part is well described by a power behaviour
$x^{\alpha}$, where $\alpha$ does not correspond to the intercept of
single $a_1$ Regge trajectory, as expected. However, the observed
power leads to the validity of Bjorken sum rule and it is consistent
with the power extracted from all previous data using NLO evolution.
At the same time, the isoscalar part behaviour may be interpreted as
a partial cancellation between a positive non-singlet contribution
and a strongly negative singlet one.
Further experimental consequences are mentioned.

\end{abstract}

\narrowtext
\section{Introduction}

The very accurate measurement of the neutron spin-dependent
structure function $g_1^n$ \cite{Hughes}, whose results have been presented
recently, possess two remarkable
properties. First, the values of neutron structure function at $Q^2=5GeV^2$
are rather large
and negative in the region of moderately low $x$. Second, the data can be
rather accurately fitted by the power function $x^{-0.8}$ and this power
seems to be unexpectedly large. It is not obvious, where this
large number, for which there is no
indication in the proton data, is coming from,
so it may affect the extrapolation to $x=0$ and cast some doubts on
the validity of Bjorken sum rule.

In the present paper we perform the isospin decomposition of the data
for both proton and neutron.
As a result, we conclude that the isovector contribution is well approximated
by the power behaviour found earlier by an elaborate method based on NLO
evolution \cite{BFR}. It may be interpreted either as the manifestation
of $ln^2x$ terms \cite{rysns} or as the contribution of the Regge cut, produced by the
$a_1$ meson and a BFKL pomeron which has a high intercept.
At the same time, for the
isoscalar part one has the
signature for a rather singular singlet contribution, compatible
with the similar behaviour predicted by QCD \cite{Ryskin}, although statistical
errors are still rather large.
This may imply a significant gluon polarization in the nucleon in this
range of $x$, which could also clearly show up in double helicity
asymmetries $A_{LL}$ at RHIC, if they turn out to be larger
than estimated earlier\cite{RHIC}.

\section{Bjorken sum rule validity}

The main feature of the new neutron data \cite{Hughes}
is large and negative $g_1^n \sim -g_1^p$,
measured with a good accuracy
up to small $x$, say $x \sim 0.01$.
The SLAC proton data \cite{p} at $Q^2=3 GeV^2$ are positive and of roughly
the same magnitude, but on the
contrary, rather flat in this region.
This could be understood, qualitatively, as a result of the
interplay of a negative contribution at low $x$, responsible for
the singular behaviour of the neutron, and a
positive contribution at larger $x$.
To check this assumption it is instructive to consider the isovector
contribution. Since there is no clear evidence of scaling violations
between $Q^2=3 GeV^2$ and $Q^2=5 GeV^2$ in any
polarized deep inelastic scattering experiment, we may neglect, for the time
being, the effects of QCD evolution,
because we are only interested in the results, provided by the data at the
present level of statistical accuracy.

It is then possible, by combining the SLAC neutron data with the
SLAC proton data, to determine the quantity $g_1^{p-n} \equiv g_1^p-g_1^n$,
entering the Bjorken sum rule, with a higher accuracy, than for proton alone.
This is because the neutron data are fortunately negative, so the
difference is larger in magnitude than $g_1^p$, while the errors are
 practically the same for $g_1^{p-n}$ and $g_1^p$, given
the better accuracy of the neutron data.

Performing this analysis of the data, we see immediately a behaviour less
sharp than that of neutron.
Since phase space effects, as powers
of $(1-x)$, are not so important in the region under consideration, we
were looking for a simple power parametrization of the data,
implied by Regge pole behaviour, namely

\begin{equation}
g_1^{p-n}(x)=
g_1^{p-n}(x_0)\left({x\over {x_0}}\right)^{-a},
\end{equation}
and we found that
this works rather well, for $0.016 \leq x \leq 0.125$ with
 the following choice of parameters:

\begin{equation}
g_1^{p-n}(x)=0.147 x^{-0.45},
\end{equation}
as shown in Fig.1.
The power we obtain is significantly smaller than
the expected contribution of the $a_1(1260)$ meson trajectory $(\sim 0.14)$.
As a result, the contribution to the Bjorken
integral from the region $0 \leq x \leq 0.125$ is large

\begin{equation}
\int_{0}^{0.125}dx g_1^{p-n}(x)=0.085.
\end{equation}

The region of higher
$x$ corresponds to a neutron contribution much smaller than that
of the proton, the latter also providing a large contribution to the integral,
and we have

 \begin{equation}
\int_{0.125}^{1}dx g_1^{p-n}(x)\approx \int_{0.125}^{1}dx g_1^{p}(x)=0.09.
\end{equation}

The total contribution to the Bjorken sum rule,

\begin{equation}
\int_0^{1}dx g_1^{p-n}(x)=0.175
\end{equation}
appears to be in the fair agreement with the theoretical value.
Note that when the final neutron data will be available, as well as more
precise proton data, it will allow one a more serious analysis, taking also
into account the effects of QCD evolution.

At the present moment we would like to stress, that by combining
the current neutron and proton data, one is led to good agreement with
Bjorken sum rule.
Let us stress that this power is compatible with the one obtained using
next-to-leading order (NLO) fit \cite{BFR} to all previous data 
$(-0.56 \pm 0.21)$. It is consistent with the result of $ln^2x$ summation
\cite{rysns}. 
We may also understand the observed power  
$\alpha=-0.45$ by considering
the intercept of the Regge cut associated to the $a_1$ meson and
a Pomeron, namely
\begin{equation}
\alpha=1-\alpha_P-\alpha_{a_1}
\end{equation}
provided we use the famous BFKL Pomeron \cite{BFKL} with the intercept
$\alpha_P \sim 1.6$ and $\alpha_{a_1}=-0.14$.

\section{Isoscalar channel and the singlet contribution}.

Since we found, that the sharp neutron structure function is not seen
in the difference between proton and neutron, it should be attributed
to the isoscalar channel. Also, the partial cancellation, we
suspect to be at the
origin of a flat proton structure function, should be manifested
in this channel as well. To check this, we calculated the quantity
$g_1^{p+n} \equiv g_1^p+g_1^n$. It really shows a
rather flat structure for $x \geq 0.035$.
The relative errors are much larger in this case because $g_1^{p+n}$
is small due to the fact that $g_1^p$ and $g_1^n$ have opposite signs.
Of course this fact also implies a small
value of the deuteron structure function in this kinematic region,
which is barely consistent with the existing data \cite{D}.

This flat structure should be related to the interplay of the negative
sharp contribution, showing itself in the fit $x^{-0.8}$ to the neutron
data \cite{Hughes},
and a positive contribution with a smaller power, dominating at larger $x$.
Since there is no counterpart for such a sharp behaviour in the conventional
Regge analysis, we make a strong, but natural assumption. Namely, we suggest
 that it is
manifested in the $SU(3)$-singlet channel.

It is in this channel
that a strong mixing between polarized quarks and gluons provides the
anomalous gluon contribution to the first moment of $g_1$ \cite{EST}.
Recent studies show that generalized
anomalous gluon contribution appears also
in all moments\cite{Muller,ST95}. At low $x$
the quark-gluon mixing
provides a strong correction to the subleading behaviour \cite{Ryskin},
producing a power close to 1. One might expect that a similar effect
is also present in the non-perturbative region where it can give rise
to a strong $x$ dependence in gluon distribution at low $Q^2$, which
is an initial condition in the approach just mentioned above.
An example of such a non-perturbative contribution is given by
instanton effects\cite{Dorohov} which,however, do not provide yet a reliable
quantitative estimate.

Moreover, the $SU(3)$-
nonsinglet part receives the contributions from the $f_1(1285)$- and
$\eta(547)$-mesons trajectories.
The first one has an intercept close to that of $a_1$ occurring in the
isovector channel,
while the second one produces a smoother behaviour like $x^{0.3}$.
For the first estimate we neglect the latter and find that the data are
well described by the formula:

\begin{equation}
g_1^{p+n}(x)=0.145 x^{-0.45}-0.03 x^{-0.87},
\end{equation}
as seen in Fig.2.
This formula is suggesting that the isoscalar contribution is approximately
equal to the isovector one, which is not so surprising, in order to have
the neutron structure function, dominated
by the most singular power only. This would mean, that in this region of $x$,
say between 0.01 and 0.1, one has

\begin{equation}
\Delta u(x)-\Delta d(x) \sim \Delta d(x) - \Delta s(x),
\end{equation}
requiring a strong negative $s$-quark polarization.
Apart from the difficulties of
incorporating this result to current models of nucleon structure, it could
also conflict with the Bjorken sum rule for the decay
of strange baryons, implying that the integrals of both sides of this
equation should be of opposite signs. Although the suggested above equality
may be valid only in a limited region of $x$, and violations
of $SU(3)$ symmetry may be  possible, it is more likely, that the $\eta$
contribution
makes the $x$ dependence of isovector and isoscalar combinations, different
in the region under consideration, so we will have

\begin{equation}
g_1^{p+n}(x)=C_{ns}x^{-0.45}+C_\eta x^{0.3}-C_s x^{-a_s}.
\end{equation}

The numerical analysis shows,
that the present inaccurate data, mainly for $x \leq 0.035$
 are equally well described
with a wide range of coefficients $C_f$ and $C_\eta$. Bearing in mind the
problem with the second Bjorken sum rule mentioned above,
 one may suspect a negative value
for $C_\eta$ in order to reduce the isoscalar integral at larger values of $x$.
On the other hand, the parameters of the singlet contribution are rather
stable, $C_s \sim 0.03, a_s \sim 1$, which is again related to the
good accuracy of neutron data.

Note that, if this neutron behaviour is really an isoscalar phenomenon,
as suggested by our analysis, one should observe a decrease and
sign change for the proton structure function not too far from
$x \sim 0.005$ at low $Q^2$.
This is the first major check of this result. However, negative proton
structure function for much smaller values of $x$ and low $Q^2$ have been
considered in the literature and a sign change may also come at large $Q^2$
from the effect of QCD evolution \cite{Blumlein}.

Note that the obtained power is also compatible with the NLO fit \cite{BFR}
result, but the accuracy of the neutron data would allow to reduce the error.

 It is, of course, too early to relate unambiguously
the observed behaviour to the results
of \cite{Ryskin}, because of the limited experimental accuracy and some
theoretical problems. In particular, it is not absolutely
clear, to what values of
$x$ and $Q^2$ the results of \cite{Ryskin} should be
 applicable. Nethertheless, the
relative closeness of the experimental and theoretical numbers may be a signal,
that the low $x$ asymptotic behaviour is manifested rather early, especially
in the neutron case, where it is not screened by a large
nonsinglet contribution, like in the proton case.
Note that clear evidence for negative $g_1^n$ no longer requires
the negative gluon polarization, as was guessed in \cite{Ryskin},
relying on earlier data.
Moreover, we found
that the formula (3.24) of \cite {Ryskin} is not incompatible
with the data, if some mean values (like suggested in \cite{Ryskin})
of parton distributions
are taken as an input. However, this approach does not allow one to
extract the $x$-dependent parton distribution, and we shall use
now the continuity
with the region of average $x$ in order to get an estimate.

It is not clear, to which extent one should attribute a small $x$ singlet
contribution to quarks or to gluons. However, according to \cite{Ryskin}
the contribution of gluons is dominant, so we neglect the quark contribution
in the present approach.
Requiring the qualitative continuity in transition
between low and average $x$, and applying the parton-like
formula for the anomalous gluon contribution it seems natural to expect
that the gluon distribution is behaving like

\begin{equation}
\Delta G(x) \sim x^{-0.87}.
\end{equation}
Note, that this formula is assuming the simple relation between
singlet contribution to $g_1$ and $\Delta G$ at average $x$
\begin{equation}
g_1^s(x)=-{\alpha_s\over {6 \pi}}\Delta G(x),
\end{equation}
which is, strictly speaking, is valid for the first moment only.
More generally, one has \cite{Karl}
\begin{equation}
g_1^s(x)=-{\alpha_s\over {6 \pi}}\int_x^1 {\Delta G(z)\over z}
E({x\over z})dz,
\end{equation}
where $E(y)$ is the coefficient function, describing the gluon-photon
interaction. Recent studies, based on the non-local generalization of
the axial anomaly\cite{Muller}, support the following choice \cite{ST95}
\begin{equation}
E(z)=2(1-z),
\end{equation}
leading, for the gluon distribution of the type $const \times x^{-a}$,
to the relation
\begin{equation}
g_1^s(x)=-{2\over {a(a+1)}}{\alpha_s\over {6 \pi}}\Delta G(x).
\end{equation}

As a result, for $a=0.87$, $\Delta G(x)$ should be multiplied, for a given
$g_1^s(x)$, by a factor $\sim 0.8$ and the integral of $\Delta G(x)$ in the
range $0.01 \leq x \leq 0.1$ is about 1.

The double helicity asymmetry $A_{LL}$ in prompt photon production
in $pp$ collisions is directly related to $\Delta G(x)$
and for the center of mass energy $\sqrt{s}=500 GeV$, which will be reached
at RHIC, one is probing precisely this kinematic region of $x$.
So given such a strong gluon polarization, we anticipate a larger $A_{LL}$
than previously predicted \cite{RHIC}. Similar comments can be made for
$A_{LL}$ in inclusive jet production.

\section{Discussion and Conclusions}

We have performed here an
analysis based on the present level of the experimental
accuracy, still not enough to see the effects of QCD evolution, but
providing interesting information about isospin structure.
For this reason we restrict ourself to the SLAC data, and neglect
the effects of QCD evolution in our analysis.
Note that the SMC data \cite{SMC} are unfortunately not accurate enough
to be used for such a simple isospin decomposition. The results on
$g_1^d$ are certainly not incompatible with our $g_1^{p+n}$ but
in order to obtain $g_1^{p-n}$, it is necessary to extract
$g_1^n$ from $g_1^d$ after subtracting $g_1^p$, which enhances
substantially the statistical errors. However, the elaborate statistical
analysis using NLO evolution lead to the very similar powers, although
the error for the singlet case is still very large.

The presented simple picture of the nucleon structure is based on the two
observations.

i) As suggested by the E154 Collaboration,
 the $g_1^n$ behaviour is well described by $\sim x^{-0.8}$.

ii) From our simultaneous analysis of proton and neutron data,
there is no indication of such a behaviour in $g_1^{p-n}$. Instead,
it is well described by the $\sim x^{-0.45}$, which leads
to a good saturation of Bjorken sum rule.

Consequently, the existence of a strong negative isoscalar contribution
is implied by these two facts. It seems rather well established,
and leads to predict a negative $g_1^p$ for $x$ below $0.005$.

Both the interpretation of the nonsinglet behaviour as a $ln^2x$ terms (or cut
produced by the BFKL pomeron), as well as
the relation of the sharp behaviour of the singlet contribution
at low $x$, and even further, to a strong gluon polarization, can be
considered more speculative.
It would be an unusual coincidence, that two rather different aspects of
small-$x$ physics manifest themselves in the same physical quantity.
However, these assumptions seem to us possible,
and they will be either supported or disproved by future
more accurate data which will allow to elaborate a better
analysis of the problem.

We are indebted to C. Bourrely, A.V. Efremov, J. Ellis, S. Forte and E. Hughes
for stimulating discussions and valuable comments. 
O.T. is grateful to Centre de Physique Th\'eorique for warm hospitality
and to the Universit\'e de Provence for financial support. He was
partially supported by Russian Foundation of Fundamental Investigation
under Grant 96-02-17361.
This investigation was supported in part by INTAS Grant 93-1180.

\newpage
\begin{figure}[ht]
\hfill
\begin{minipage}{6.5in}
\label{fig1}
\caption{Comparison for $g_1^{p-n}$ between the curve given by eq. (2)
and the SLAC data refs.\protect\cite{Hughes,p}. For $g_1^p$ at $x=0.0165$
due to the absence of SLAC data we used the SMC data ref.
\protect\cite{SMC}.} 
\end{minipage}
\end{figure}

\begin{figure}[ht]
\hfill
\begin{minipage}{6.5in}
\label{fig2}
\caption{Same as Fig.1 for $g_1^{p+n}$
with the curve given by eq. (7) }
\end{minipage}
\end{figure}

\end{document}